\documentstyle[12pt]{article}

\newcommand{\be}{\begin{equation}}
\newcommand{\ee}{\end{equation}}
\newcommand{\ba}{\begin{eqnarray}}
\newcommand{\ea}{\end{eqnarray}}
\newcommand{\no}{\nonumber \\}

\begin{document}

\begin{center}

{\Large Electric charge and magnetic flux on rotating black holes
  in a force-free magnetosphere}

  \vskip 1cm

  Hyun Kyu
  Lee$^{a,c,}$\footnote{e-mail: hklee@hepth.hanyang.ac.kr}, Chul H.
  Lee$^{a,}$\footnote{e-mail: chlee@hepth.hanyang.ac.kr}
  and Maurice H.P.M. van Putten$^{b,c,}$\footnote{e-mail:
  mvp@math.mit.edu}

  \vskip .5cm

  $^a$ Department  of Physics, Hanyang University, Seoul 133-791,
  Korea\\ $^b$ MIT, Cambridge, MA 02139, USA \\ $^c$ Department  of
  Physics and Astronomy, State University of New York, Stony Brook,
  NY 11794-3800, USA
  \end{center}

\newpage

\noindent {\large Abstract}

     The electric charge on rotating black holes is
  calculated to be $\sim BJ$ in the force-free configuration of
  Ghosh (2000), with a horizon flux of $\sim BM^2$. This charge is
  gravitationally weak for $B\sim 10^{15}$G, so that the Kerr metric
  applies. Being similar to the electric charge of a magnetar, both
  electric charge and magnetic flux should be, in sign and order of
  magnitude, continuous during stellar collapse into a black hole.
  Extraction of the rotational energy from newly formed black holes
  may proceed by interaction with the magnetic field.\\ \mbox{}

\section{Introduction}

     The rotational energy of a central supermassive black hole
  is potentially important in powering active galactic nuclei (AGN),
  in addition to the energy liberated in accretion (Lynden-Bell
  1969, Lynden-Bell and Rees 1971).  The accretion flow is probably
  magnetized, which  facilitates the extraction of energy of the
  accretion flow(Lovelace 1976) and, possibly, that of the
  rotational energy of the black hole as well (Blandford and Znajek
  1977). These ideas have recently been re-examined in the context
  of gamma-ray bursts, where a low mass black hole is threaded by a
  magnetic flux of similar magnitude (Meszaros and Rees 1997;
  Paczynski 1998; van Putten 1999, 2000; Lee, Wijers and Brown
  2000).

     In the Blandford-Znajek process, a rotating black hole
  interacts with the surrounding magnetosphere in the force-free limit.
  The electric field on the horizon has a non-vanishing normal component,
  which may be interpreted in terms of a surface density of electric charge
  in accord with the membrane paradigm developed by
  Thorne, Price and Macdonald (1986). The integral of the surface
  charge defines a net electric charge $Q$. All ``three hairs"
  of mass, $M$, angular momentum, $J$, and $Q$, therefore,
  appear in black hole-magnetic field interactions.

     It is well-known that rotating black holes in an
  aligned   vacuum magnetic field assume a horizon charge $Q=2BJ$ in
  their lowest energy state (Wald 1974, Dokuchaev 1987). Here, $J$
  denotes the angular momentum and $B$ the asymptotic
   magnetic field-strength. Since the horizon charge corotates
  with the horizon, it induces a magnetic flux (van Putten, 2000)
  \begin{eqnarray}
  \phi_H=4\pi QM\Omega_H=8\pi BM^2\sin^2(\lambda/2)
  \end{eqnarray}
  where $\sin\lambda=a/M$ and $\Omega_H$ parametrizes the angular
  velocity of the black hole (van Putten 1999). Together with the
  charge-free flux, this recovers a net horizon flux (Wald 1974,
  Dokuchaev 1987, van Putten 2000)
   \begin{eqnarray}
  \Phi_H=4\pi BM^2 \cos\lambda+\phi_H=4\pi BM^2
  \end{eqnarray}
  - an electrostatic equilibrium  in which the metric is the Kerr
  metric (Dokuchaev 1987, Kim, Lee and Lee 2000).

     This observation suggests the possibility that a magnetic flux on
  a rapidly rotating black hole is similarly maintained by a finite
  horizon charge also when the magnetosphere is force-free. It may be
  noted that the electric horizon charge has been calculated for inflow
  in the approximation of ideal magnetohydrodynamics (Ruffini \& Wilson,
  1975). The force-free limit considered here is different, as a singular
  limit defined by neglecting Reynolds stresses. The presented calculation
  is of particular interest to the high-angular momentum black hole-disk
  or torus model of gamma-ray bursts with magnetic fields $B\sim 10^{15}$G
  (see, e.g., van Putten \& Ostriker (2000) for a recent discussion),

     In \S2, we calculate the electric charge on the
  black hole in a force-free magnetosphere given by Ghosh (2000).
  The result $Q\sim BJ$ is gravitationally weak, wherein the Kerr
  metric remains a valid approximation. This result is found to be
  similar to the charge on a magnetar in \S3. The electric charge
  and flux of the core of a star in collapse, therefore, is
  continuous in order of magnitude. The astrophysical aspects are
  summarized in \S4.

  \section{Horizon charge in a force-free magnetosphere}

  Coulomb's law of electrodynamics

  \be
  \nabla \cdot {\bf E} = 4\pi \rho_{e}, \label{coul}
  \ee
  suggests that the net electric charge $Q$ can be obtained
  by the integral
  \be
  Q = \frac{1}{4\pi} \oint d{\bf S} \cdot {\bf E} \label{charge}
  \ee
  over a closed surface surrounding the system at hand. These equations hold
  not only in flat space-time but can also be shown to hold in
  curved space-time, provided that the dot product is interpreted covariantly with
  appropriate definitions of the electric field ${\bf E}$ and the
  charge density $\rho_{e}$. This has been
  made explicit by Thorne and Macdonald (1982) in the 3+1
   formulation of electrodynamics in curved space-time.
   At each space-time point, a fiducial reference frame is chosen by
   splitting the 4-dimensional space-time into three space directions
     and one ``universal time" direction. Thus,
     Coulomb's law assumes similar expressions in curved and flat
     space-time.

     By Eq.(\ref{charge}), we can determine the charge on a
     black hole, given the electromagnetic field
     surrounding the black hole. The force-free magnetosphere in the Kerr
     metric is potentially a realistic example, as in, e.g., Okamoto (1992).
     This applies when the magnetic energy density is not large
     enough to affect the background metric significantly: $B^2M^2<<1$.

     The axially symmetric force-free magnetosphere surrounding a Kerr
     black hole is described by surfaces of continuous magnetic flux
     $\Psi(r,\theta)$ in Boyer-Lindquist coordinates
     $(r,\theta,\phi)$. These surfaces are in rigid rotation
     (Thorne, Price and Mcdonald 1986) with angular velocity
     $\Omega_F=\Omega_F(\Psi)$. The poloidal parts of  the magnetic and
     electric field  are  given by

     \be
      {\bf B}^P = \frac{\nabla \Psi \times {\bf e}_{\hat{\phi}}}{2\pi \varpi}
      , \,\,\, {\bf E} = -{\bf v}_F \times {\bf B}^P
      \label{efield}
     \ee
     where $\varpi=({\Sigma}/{\rho}) \sin \theta$ and
     $\omega={2Mra}/{\Sigma^{2}}$ is the angular velocity of the black hole.
     Also, $\alpha={\rho
     \sqrt{\Delta}}/{\Sigma}$ denotes the lapse function and ${\bf v}_F=
     {\alpha}^{-1}(\Omega_F -\omega) \varpi \, {\bf
     e}_{\hat{\phi}}$ denotes the velocity of the rotating poloidal field
     line relative to the ZAMO. Here, we use the
     Boyer-Lindquist expressions
     $\Delta = r^2 + a^2 -2Mr, \,\, \rho^2 = r^2 + a^2 \cos^2 \theta,
     \,\,\, \Sigma^2 = (r^2 + a^2)^2 - a^2 \Delta \sin^2\theta$.

     The components of the electromagnetic field
     normal to the horizon are
     \ba
     E_n &=& -\frac{\Omega_F -\Omega_H}{2\pi \alpha}
     \frac{\sqrt{\Delta}}{\rho} \, \partial_r\Psi \\
      B_n &=&
     \frac{1}{2\pi\varpi\rho} \, \partial_{\theta} \Psi \ea It may be
     noted that  there are also non-vanishing $\theta$-components of
     electric and magnetic field in a force-free magnetosphere. This
     permits ${\bf E}\cdot {\bf B}=0$ to be satisfied, even within the
     inner light surface of Znajek (1977) down to the horizon, in the
     presence of electric currents. The the total electric
     charge on the stretched horizon is given by

     \ba
     Q = -\int_{horizon}(\Omega_F -\Omega_H)\frac{\Sigma^2}
           {4\pi\rho^2} \partial_r \Psi \, \sin \theta \, d\theta, \label{Q}
     \ea
      and the total magnetic flux through upper hemisphere by
     \be
     \Phi_B \,  = \int_0^{\pi/2} B_n 2 \pi \varpi \rho \, d\theta =
     \Psi(\theta=\pi/2) - \Psi(\theta = 0).
     \ee
     $\Omega_F(\Psi)$, the electric current flowing into the black hole $I(\Psi)$,
     and $\Psi$  are determined by
     the Maxwell equation for the force-free magnetosphere, which leads
     to the equation(Blandford and Znajek 1977, Macdonald and Thorne
     1982, Beskin 1997) given by the Grad-Shavranov (GS) equation
     \ba
     \nabla\cdot\left\{{\alpha\over\varpi^2}\left[1 - {(\Omega_F -
     \omega)^2\varpi^2\over\alpha^2}\right]\nabla\Psi\right\} &+&
     {\Omega_F - \omega\over\alpha}{d\Omega_F\over d\Psi}(\nabla\psi)^2
     \no && +{16\pi^2\over\alpha\varpi^2}I \frac{dI}{d\Psi} =
     0.\label{stream}
      \ea

      Recently Beskin and Kuznetsova (2000) showed by solving the GS
      equation that for transonic flow onto a black hole the poloidal
      magnetic field in
     the supersonic domain near horizon remains actually the same as
     one can find from the force-free solution.  It implies that in
     calculating the electric charge, Eq.(\ref{Q}),  one can use the
     force-free poloidal magnetic field obtained from
     Eq.(\ref{stream}).  This stream equation is quite complicated and
     does not appear to allow exact analytical solutions.

     In the restricted case of a non-rotating black hole with $\Omega_F = 0$,
     Eq. (\ref{stream})  reduces to the following simple form;
     \be
     \nabla\cdot\left\{{\alpha\over\varpi^2}\nabla\Psi\right\}
     =0.\label{stream0}
     \ee
     General properties of Eq. (\ref{stream0}) are
     discussed by Ghosh (2000). Although we do not have the solution
     $\Psi$ in the general rotating black hole case, a numerical
     study by Macdonald (1984) demonstrates that the effect of the
     black hole rotation to the poloidal magnetic field structure is
     small even at high specific angular momentum $a \leq 0.75M$. Based on
     these observations we will assume in this work that the poloidal
     structure of the magnetosphere surrounding a black hole in the rotation
     case can be inferred from the nonrotating case.
     Hence, following Ghosh and Abramowicz (1997), we have
     for the upper hemisphere,
     \be
     \Psi = \Psi_0[(r-2M)(1-\cos\theta) - r_0\cos\theta -
     2M(1+\cos\theta)\ln(1+\cos\theta)]\, \label{Psi1}
     \ee
     and for the lower hemisphere,
     \be
     \Psi = \Psi_0[(r-2M)(1+\cos\theta) + r_0\cos\theta -
     2M(1-\cos\theta)\ln(1-\cos\theta)]\, \label{Psi2}
     \ee
     which are linear combinations of the specific solutions discussed
     by Macdonald (1984) and Ghosh (2000). This naturally represents
     the situation where an accretion disk exists on the equatorial
     plane with the toroidal component of the current density
     \begin{equation}
      j^{\hat{\phi}}=\frac{\Psi_{0}(r+r_{0})}{4\pi^{2}r\Sigma}
     \delta(\theta-\frac{\pi
     }{2}).
     \end{equation}

     The calculation of the electric charge on the black hole is now
     given by Eqs. (\ref{Q}), (\ref{Psi1}), and
     assuming an optimal  case where $\Omega_F \sim \Omega_H/2$, we have
     \ba
     Q =\frac{\Psi_0}{\pi}\frac{J}{r_H}[\frac{r_H}{a} \tan^{-1} \frac{a}{r_H}
     -\frac{1}{2}(\frac{r_H}{a})^2 \ln \{1 + (\frac{a}{r_H})^2\}].
     \ea
     In using an approximate solution for obtaining an order of magnitude
     estimate, we consider
     $\Psi_0 \sim \frac{\Phi_B}{4M}$ and $ \Phi_B \sim 8\pi M r_H B$ to find
     \ba
     Q \sim \frac{1}{8\pi} \frac{J\Phi_B}{M r_H}= BJ\label{qjb}
     \ea
     This result is similar to the Wald charge in the ground state of the
     black hole in a vacuum magnetic field. For a rapidly rotating black hole
      with strong magnetic fields ($B =10^{15}$G), in the context of GRBs,
      the charge can be estimated as
     \ba
     Q \sim 10^{15} C \left( \frac{B}{10^{15}G} \right) \left( \frac{M}{M_{\odot}} \right)^2
     .\label{qjbz} \ea
      Compared to the extremal charge on the
     black hole, one can see that  the above charge is much smaller
     \be
     Q \ll Q_{extremal} = 4.2 \times 10^{20} C \left( \frac{M}{M_{\odot}} \right)
     \ee
     and therefore the Kerr metric can be treated safely as a
     background for the force-free magnetosphere.  One can see also that
     this electric charge can not provide
     a source of significant energy for a stellar mass black hole, when
     compared to the rotational energy.

     \section{Magnetic flux on black holes and magnetars}

     Pulsars are known to be rapidly rotating neutron stars with
     strong magnetic fields (Shapiro \& Tuekolsky 1983). Recently,
     several magnetars with even stronger magnetic fields (Kouveliotou et
     al. 1999) have been
     found. The vacuum surrounding these compact objects is unstable which
     evolves into a magnetoshpere with charged particles.
     Goldreich-Julian (Goldreich \& Julian 1969) suggested a model of
     a degenerate pulsar magnetosphere in the force-free limit.
     Assuming the poloidal field structure to be a magnetic dipole
     \be
   {\bf B}^P = \frac{B_0 R^3}{r^3} (\cos\theta \, {\bf e}_{\hat{r}}+
      \sin\theta \, {\bf e}_{\hat{\theta}}) \label{nbp}
    \ee
       the electric field using the degenerate condition becomes
   \ba
       {\bf E} = - {\bf v}_F \times {\bf B}^P, \,\,
    {\bf v}_F = {\bf \Omega} \times {\bf r} \label{vb} \ea where ${\bf
       v}_F$ is the velocity  of the rigidly rotating field lines.
   The surface charge density and the electric charge on the rotating
   neutron star in a degenerate magnetosphere (Goldreich and Julian 1969,
   Ruderman and Sutherland 1975) becomes
     \be
     Q_{NS} = \frac{B_0 \Omega R^3}{3} \label{qns}
     \ee
     Using the total magnetic flux (the stream function on the neutron
     star surface reads $\Psi(\theta) = \Phi_B(1- \cos^2 \theta)$)
     through upper hemisphere,   $\Phi^{NS}_B = \pi R^2 B_0$, and
     assuming the neutron star as a rigid sphere,
     $J_{NS} =({2}/{5})MR^2\Omega $,   we have
     \ba
     Q_{NS} = \frac{5}{6\pi} \left(\frac{J\Phi_B}{MR}\right)_{NS}
     \label{QNS}.
     \ea
     Thus, the same structure as in Eqn. (\ref{qjb}) is obtained,
     both in sign and order of magnitude. Taking
     $M\sim 1.5 M_{\odot}, \,\,\, R \sim 10^{6} \mbox{cm}, \,\,\, \Omega \sim
     10^{4}/$s for a magnetar with $B \sim 10^{14}$G, we conclude
     $Q_{NS} =10^{15}$C - the same sign and order of magnitude as that
     on the rotating black hole in a force-free magnetosphere obtained
     above.

     These results give an indication of the charge during accretion-induced
     or prompt collapse into a black hole. During collapse,
     the specific angular momentum $a=J/M$ will
     change continuously. Since the time-scale of accretion of a charge $BJ$
     is of order $M$, the notion that both the neutron star and the final
     black hole state have charges of order $\sim BJ$ indicates that the
     charge $Q$ should remain
     about this value. This would hold in particular when the accreting matter
     is magnetized at the level of the collapsing neutron star, in which case
     the initial charge on the neutron star and the final charge on the
     black hole are similar within a factor of unity. In this event,
     with $r_H/R_{NS} \sim  10^{-1}$ we get
     from Eq. (\ref{qjb}) and Eq. (\ref{QNS})
     \ba
     \Phi_{BH} \sim \Phi_{NS}.
     \label{flux}
     \ea

  The continuity of magnetic flux provides the necessary
  horizon flux on the black hole  for tapping the rotational
  energy, for example, via the Blandford-Znajek process. The
  relationship (16) allows
  the power in Poynting flux to be alternatively written as
  \ba
  P_{BZ} \sim 10^{50} \mbox{erg/s} \left(\frac{aB}{10^{-4}} \right)^2
          = 10^{50} \mbox{erg/s} \left(\frac{Q/M}{10^{-4}} \right)^2.
  \ea
  The expression on the right brings about the relationship between $P_{BZ}$
  and the associated charge-to-mass ratio. The charge-to-mass
  ratio remains small for stellar mass black holes in astrophysical environments,
  which is consistent with the assumed Kerr geometry.

     \section{Summary}

     The electric charge on a rotating black hole immersed in a
     force-free magnetosphere is estimated to satisfy $Q\sim BJ$ in
     the poloidal structure of Ghosh and Abramowicz (1997). For
     canonical GRB/magnetar values $B\sim 10^{15}$G, the gravitational
     contribution of the electromagnetic field can be neglected.
     Also black holes and magnetars are found to carry similar electric
     charge in sign and magnitude.

     We interpret these results to indicate that the central charge
     in the collapsar/hypernova scenario of GRBs remains essentially
     continuous and, with it, the associated magnetic flux.
     Provided that the magnetic field continues to be supported by
     surrounding magnetized matter, e.g., fall back matter stalled
     against an angular momentum barrier, extraction of rotational
     energy from the newly formed black hole may continue by interaction
     with the magnetic field. Wherever the magnetosphere is force-free,
     the interaction with the magnetic field is
     notably so in the form of a Poynting flux
     (Blandford \& Znajek, 1977; Lee, Wijers \& Brown, 2000).
     The horizon charge is important in preserving
     magnetic flux, especially when the black hole spins rapidly,
    but will not provide a source of significant energy for a
    stellar mass black hole.

     \vskip .5cm

     {\bf Acknowledgement.}
      We thank G.E. Brown for stimulating discussions and his
  hospitality at SUNY Stony Brook, where a part of this work was
     performed. We are also grateful to V. Beskin and P. Ghosh for
     valuable discussions and suggestions. HKL and CHL are supported in
     part by the interdisciplinary Research program of the KOSEF, Grant
     No. 1999-2-003-5, MVP acknowledges support from NASA Grant No.
     5-7012 and an MIT C.E. Reed Award. HKL also acknowledges supports
     from Asia Pacific Center for Theoretical Physics.

     \newpage

\noindent {\large References}

\vskip 1cm

\noindent Beskin, V. S. 1997, Usp. Fiz. Nauk, 167, 689, Eng.
Trans. in Physics - Uspekhi, 40, 659

\vskip .3cm

\noindent Beskin, V. S. Kuznetsova, I. V. 2000, Nuovo Cimento,
115, 795

\vskip .3cm

\noindent Blandford, R.D. and Znajek, R.L. 1977, MNRAS, 179, 433

\vskip .3cm

\noindent Dokuchaev, V.I. 1986, JETP, 65, 1079

\vskip .3cm

 \noindent Ghosh P. and  Abramowicz M.A. 1997, MNRAS,
292, 887

\vskip .3cm

\noindent Ghosh, P. 2000, MNRAS 315, 89

\vskip .3cm

\noindent Goldreich, P. and Julian, W. 1969, ApJ, 157, 869

\vskip .3cm

\noindent Kouveliotou, C. et al, ApJ, 510, L115

\vskip .3cm

\noindent Kim, H., Lee, C.-H., Lee, H.K. 2000,  to appear in Phys.
Rev. D,  gr-qc/0011044

\vskip .3cm

\noindent Lee, H.K.,  Wijers, R.A.M.J.,  and  Brown, G.E. 2000,
 Phys. Rep., 325, 83

\vskip .3cm

\noindent Lovelace, R.V.E. 1976, Nature, 262, 649

\vskip .3cm

\noindent Lynden-Bell, D. 1969, Nature, 223, 690; Lynden-Bell, D.,
and Rees,  M.J.  1971, MNRAS, 152, 461

\vskip .3cm

\noindent Macdonald, D. and Thorne, K.S. 1982, MNRAS, 198, 345

\vskip .3cm

\noindent Macdonald, D. 1984, MNRAS, 211, 313

\vskip .3cm

\noindent Meszaros, P. and Rees, M.J. 1997, ApJ, 482, L29

\vskip .3cm

\noindent Okamoto, I. 1992, MNRAS, 294, 192

\vskip .3cm

\noindent Paczynsky, B. 1998, ApJ, 494, L45

\vskip .3cm

\noindent Ruderman M.A. and Sutherland P.G. 1975, ApJ, 196, 51

\vskip .3cm

\noindent Ruffini, R. and Wilson, J. R. 1975, Phys. Rev. D12, 2959

\vskip .3cm

\noindent Shapiro, S. and Teukolsky, S. 1983,``Black Holes, White
Dwarfs and Neutron Stars" (John Wiley \& Sons)

\vskip .3cm

\noindent Thorne, K. S., Macdonald, D. 1982, MNRAS, 198, 339 (TM)

\vskip .3cm

\noindent Thorne, K.S., Price, R.H.,  and Macdonald, D.A. 1986,
Black Holes; ``The Membrane Paradigm" (Yale University Press, New
Haven and London)

\vskip .3cm

\noindent van Putten, M.H.P.M., Science, 284, 115 (1999).

\vskip .3cm

\noindent van Putten, M.H.P.M., Phys. Rev. Lett., 84(17), 3752
(2000).

\vskip .3cm

\noindent van Putten,  M.H.P.M. and Ostriker, E. 2001,
   astro-ph/0010440

\vskip .3cm

 \noindent Wald, R.M. 1974, Phys. Rev., D10, 1680

\vskip .3cm

\noindent  Znajek, R.L. 1977, MNRAS  179, 457

     \end{document}